\documentstyle[aps,preprint]{revtex}
\tighten
\begin{document}
\draft
\hsize\textwidth\columnwidth\hsize\csname @twocolumnfalse\endcsname

\title{Symmetry Properties on Magnetization in the Hubbard Model at Finite
Temperatures}
\author{Gang Su$^{\ast}$ and Masuo Suzuki$^{\dag}$}
\address{ Department of Applied Physics, Faculty of Science,
 Science University of Tokyo\\
1-3, Kagurazaka, Shinjuku-ku, Tokyo 162, Japan}

\maketitle

\begin{abstract}
By making use of some symmetry properties of the relevant Hamiltonian,
two fundamental relations between
the ferromagnetic magnetization and a spin correlation
function are derived for the
$d (=1,2,3)$-dimensional Hubbard model at finite temperatures.
These can be viewed as a kind of Ward-Takahashi identities.
The properties of the magnetization as a function of the applied
field  are discussed.
The results thus obtained hold true for both
repulsive and attractive on-site Coulomb interactions, and for
arbitrary electron fillings.  \\

\end{abstract}



The Hubbard model was originally proposed for interpreting the origin
of itinerant ferromagnetism in transition metals within
the framework of mean-field approxmations more than thirty years
ago\cite{hubb1,gutz,kana}.
It is now widely thought to be the simplest model for correlated electron
systems. This model is exactly solved using the Bethe ansatz in one
dimension\cite{l-w}, but the exact results for dimensions $d >1$ are sparse
(see, e.g. Ref.\cite{lieb0} for a review).

Among those rigorous results favouring the existence of
ferromagnetism, the Nagaoka-Thouless's celebrated
theorem\cite{nagao} is the first example to show the itinerant electron
ferromagnetism in a special limit (with an infinite repulsive Coulomb
interaction and one hole) in the ground state. Later, Lieb\cite{lieb}
presented another example for ferrimagnetism in this model with finite
Coulomb repulsion at
half-filling on asymmetrical bipartite lattices. Mielke and
Tasaki\cite{mietasa} showed the existence of
ferromagnetism in a class
of Hubbard models with somewhat artificial flat-band systems on decorated
lattices. Very recently M\"uller-Hartmann\cite{muller} proposed  an
idea which may open a low density
 route towards the understanding of ferromagnetism in the
Hubbard models on non-bipartite lattices.

At finite temperatures, only few rigorous results are known so far.
They seem to be disfavourable of ferromagnetism in this model.
 For instance, the absence of
magnetic orderings
was rigorously proved in one and two dimensions\cite{ghosh},
while the possibility
of magnetic long-range order (LRO) in the attractive Hubbard model was
precluded in arbitrary dimensions\cite{kubo}, etc.
As the rigorous results are rare in this case, more efforts
should be devoted. On the other hand,
since the Hubbard model
possesses a few nontrivial symmetries, which could induce some
physically important properties,  it is quite interesting to explore
them so that a better understanding on the underlying physics behind
these symmetries in this model could be achieved.

It is the purpose of this paper that,
by means of some symmetry properties of the relevant Hamiltonian,
two fundamental relations between
the ferromagnetic magnetization and a spin correlation
function are derived for the
$d (=1,2,3)$-dimensional Hubbard model at finite temperatures.
These can be viewed as a kind of Ward-Takahashi identities.
The properties of the magnetization as a function of the applied
field  are discussed.
The results thus obtained hold true for both
repulsive and attractive on-site Coulomb interactions, and for
arbitrary electron fillings. A special solution is analyzed as
an example.



We start from the Hubbard model in the presence of an
external applied field $h$ ($>0$) on a $d (=1,2,3)$-dimensional
lattice $\Lambda$ with $|\Lambda|$ sites.
The Hamiltonian reads
\begin{eqnarray}
 H =  -  \sum_{x,y \in \Lambda, \sigma} t_{xy} ( c_{x\sigma}^{\dag}c_{y\sigma}
+ H.c.) + \sum_{x\in \Lambda} U_{x} n_{x\uparrow} n_{x\downarrow}
 -  h \sum_{x \in \Lambda} (n_{x\uparrow} - n_{x\downarrow}),
\label{hamil}
\end{eqnarray}
where $c_{x\sigma}^{\dag}$ ($c_{x\sigma}$) is the creation
(annihilation) operator of an electron
at site $x \in \Lambda$ with spin $\sigma (= \uparrow, \downarrow)$,
$U_{x}$ is the on-site electron-electron interaction at $x$, and
$n_{x\sigma} = c_{x\sigma}^{\dag}c_{x\sigma}$ is the number operator of
electrons.
The elements of the hopping matrix, $t_{xy}$, satisfying $t_{xy} = t^{*}_{yx}$,
survive for the short-range hoppings of electrons in practice.
Note that $t_{xx}=0$. The total number of electrons of the system, say $N =
\sum_{x \in \Lambda, \sigma} n_{x\sigma}$, is
conserved. We have taken the Land\'e factor of electrons $g$=2, and $\mu_B =1$.
The foregoing discussion is
independent of the sign of $U_x$, and sustains valid only
for finite temperatures
which implies that one can
not extract any useful information for zero-temperature case from
this study.

Define the spin operators as
$ S^{+} = \sum_{x\in \Lambda}c_{x\uparrow}^{\dag}c_{x\downarrow}$,
$S^{-}   = (S^+)^{\dag}$,  and $
 S^{z} = \frac12 \sum_{x\in \Lambda} (n_{x\uparrow} - n_{x\downarrow})$.
They satisfy the usual SU(2) Lie algebra: $[S^{+}, S^{-}] = 2 S^{z}$,
and
$[S^{\pm}, S^{z}] = \mp S^{\pm}$. The ferromagnetic magnetization per
site is defined as
\begin{eqnarray}
m (h, T) = \frac{1}{|\Lambda|} \langle S^{z} \rangle ,
\label{magn}
\end{eqnarray}
where $\langle \cdots  \rangle = Tr [\exp (-\beta
H)\cdots]/Tr[\exp (-\beta H)]$ is the thermal average in the
canonical ensemble,
with the inverse temperature $\beta = 1/T$ ($k_B =1$).

Using the
commutator $[S^{\pm}, H] =  \pm 2 h S^{\pm}$,
and the
property of cyclicity under the trace, we obtain the following basic equation:
\begin{eqnarray}
{\cal G}(h, T) ( 1 - e^{- 2 \beta h}) = 2 m(h, T),
\label{g-m}
\end{eqnarray}
where ${\cal G}(h, T) = (1/|\Lambda|) \langle S^{+}S^{-} \rangle_{h,
T}$ is the spin correlation function.
This holds generally when the Hamiltonian without a Zeeman term satisfies O(3)
symmetry in spin space. This relates an odd spin correlation such as the
magnetization to an even spin correlation. Thus, this can be viewed as a kind
of Ward-Takahashi identities\cite{ward} in quantum field theory. It should be
noted that there are an infinite number of correlation identities\cite{suzuki}
in classical systems.
It is evident that if we know the form of ${\cal G}(h, T)$, then we
can
get $m(h, T)$ from Eq.(\ref{g-m}). On the basis of this strategy, we now
try to exploit
such properties of the function ${\cal G}(h, T)$ that it enables us
to obtain an exact solution for ${\cal G}(h, T)$.
Invoking the spin up-down symmetry which exchanges the up
spins and the down spins of electrons,
\begin{eqnarray}
 c_{x\uparrow}\rightarrow c_{x\downarrow}, ~~~
 c_{x\uparrow}^{\dag} \rightarrow c_{x\downarrow}^{\dag}
\label{u-d}
\end{eqnarray}
we find $H \rightarrow {\tilde H} = H_0 + 2 h S^{z}$ and $S^{+}S^{-}
\rightarrow S^{-}S^{+}$, where we have denoted the Hamiltonian
[Eq.(\ref{hamil})] by $H_0 - 2 h S^z$.
An alternative closed form for this unitary transformation can be found
in Ref.\cite{su1}. It is interesting to note that
this transformation, unlike the
usual particle-hole transformation, conserves the total number of
electrons (while it conserves neither the number of electrons
with up spins nor the number of electrons with down spins).
It is this property that makes it possible for us to work
in the canonical ensemble.
By applying the transformation, Eq.(\ref{u-d}), to ${\cal G}(h, T)$,
we obtain
\begin{eqnarray}
{\cal G}(h, T)  = {\cal G}(-h, T) e^{2 \beta h},
\label{g-g}
\end{eqnarray}
where we have utilized the cyclicity under the trace in ${\cal G}(h, T)$,
$S^{-}S^{+} = S^{+}S^{-} - 2 S^{z}$, and Eq.(\ref{g-m}) as well. This
equation is of basic importance for discussing the poperties  of the
correlation function ${\cal G}(h, T)$.


Generally speaking, Eq. (\ref{g-g}) has an infinite number of
solutions, but the number of the physically meaningful solutions,
which could depend on the details of the Hamiltonian of the system,
might be limited. Considering this point, we shall in the following
only search for those solutions which are physically sound.
Obviously, it would not be possible to solve
Eq. (\ref{g-g}) without first obtaining the conditions satisfied by the function
${\cal G}(h, T)$. For this purpose we differentiate ${\cal G}(h, T)$
with respect to $h$, and using the unitary transformation (\ref{u-d})
to the thermal averages involved, and then noting Eqs. (\ref{g-m})
and (\ref{g-g}), we obtain
\begin{eqnarray}
\frac{\partial {\cal G}(h, T)}{\partial h} \geq
\frac{\partial {\cal G}(- h, T) }{\partial h},
\label{g-inequ1}
\end{eqnarray}
where we have utilized the facts $[ \langle (S^{z})^2 \rangle -
\langle S^{z} \rangle^2 ]_{h,T} = [ \langle (S^{z})^2 \rangle -
 \langle S^{z} \rangle^2 ]_{-h, T}$ and
 $\langle (S^{z})^2 \rangle -  \langle S^{z} \rangle^2  \geq 0$.
In addition, by Eqs. (\ref{g-m}), (\ref{g-g}) and (\ref{g-inequ1})
we get the following inequalities:
\begin{eqnarray}
0 < {\cal G}(h, T) \leq \frac{\rho}{1-e^{- 2\beta h}}
\label{g-inequ2}
\end{eqnarray}
and
\begin{eqnarray}
\frac{\partial}{\partial h}\log {\cal G}(h, T) \geq
\frac{- 2\beta e^{-2\beta h}}{1-e^{-2\beta h}},
\label{g-inequ3}
\end{eqnarray}
where $\rho = N/|\Lambda|$ is
the density of electrons in the system.
On the other hand, we can obtain, after performing the transformation
(\ref{u-d}) to $m(h,T)$, the equation
\begin{eqnarray}
m (h, T) = - m (-h, T),
\label{m-m}
\end{eqnarray}
which implies that $m (h, T)$ must be an odd function of $h$ at
given $T$.

Up to now we have had some basic information about the function
${\cal G}(h, T)$,
namely, Eqs.(\ref{g-m}) and (\ref{g-g}) - (\ref{m-m}) should be
satisfied simultaneously. Under these conditions, we can solve
Eq. (\ref{g-g}) exactly. The result, being surprisingly simple,
takes the following form:
\begin{eqnarray}
{\cal G}(h, T) = \frac{f(h,\rho,T)}{1 + e^{- 2\beta h}},
\label{g-solu}
\end{eqnarray}
where the function $f(h,\rho,T)$, whose properties are to be discussed below,
is an even function of $h$.
Substituting this solution into Eq. (\ref{g-m}) one gets
\begin{eqnarray}
m(h, T) = \frac{f(h,\rho,T)}{2} \tanh (\beta h).
\label{m-f}
\end{eqnarray}
 Evidently,  Eqs. (\ref{g-solu}) and (\ref{m-f})
satisfy Eq. (\ref{g-g}) and Eq. (\ref{m-m}), respectively.
  Eq.(\ref{m-f}) should remain valid in
the thermodynamic limit ($N \rightarrow \infty$ and $|\Lambda| \rightarrow
\infty$ with the ratio $N/|\Lambda| = \rho$ fixed). As a
consequence,  the  magnetic susceptibility can be obtained by $
\chi (T) = \frac{\partial m(h,T)}{\partial h}|_{h \rightarrow 0}$.
Eq.(\ref{m-f}) forms the main result of this paper, which is somewhat
general for the Hubbard model defined in Eq.(\ref{hamil}).

Now, we are
in the position to discuss the properties of the function $f(h,\rho,T)$.

1. As mentioned before, $f(h,\rho,T)$ is an even function of the
applied field, namely,
\begin{eqnarray}
f(h,\rho,T) = f(-h,\rho,T).
\label{f-even}
\end{eqnarray}

2. From Eqs. (\ref{g-inequ2}) and (\ref{g-inequ3}), one may find that
for any values of $h$ the function $f(h,\rho,T)$ should
satisfy the following two inequalities:
\begin{eqnarray}
& & 0 < f(h,\rho,T) \leq \rho \coth (\beta h), \\
\label{f-inequ1}
& & \frac{\partial}{\partial h} \log f(h,\rho,T) \geq -
\frac{2\beta }{\sinh (2\beta h)}.
\label{f-inequ2}
\end{eqnarray}
(Recall that $h>0$.)

3. From the physical
point of view, we note that $m \rightarrow m_{max} \equiv S_{tot}/|\Lambda|$
as $h \rightarrow + \infty$ at given $T$,
because in this limit all spins of electrons should
align on the direction of the applied field, where $S_{tot}$ is the total
spin of the system, i.e., $S_{tot} = N/2$ in our case. $m_{max}$ is
therefore the saturation magnetization per site. On account of this
consideration, we find
\begin{eqnarray}
\lim_{h \rightarrow \infty} f(h,\rho,T) = \rho.
\label{f-infty}
\end{eqnarray}
This constraint is a consequence of the well-known physical result:
$|m(h,T)| \leq m_{max}$ for any $h$ and finite $T$.

4. S\"ut\H o \cite{suto} proved an
asymptotic expression for the ferromagnetic magnetization per site in
the Hubbard model:
$m(T,h,\rho) \rightarrow \tanh(\beta h)$ as $\rho \rightarrow 1$ (note
that there is a difference in the definition of the magnetization
by a prefactor $1/2$). By means of this result, we have
\begin{eqnarray}
\lim_{\rho \rightarrow 1} f(h,\rho,T) = 1.
\label{f=1}
\end{eqnarray}

5. The function $f(h,\rho,T)$ might be dimension-dependent, i.e., it
may probably have different forms in different spatial dimensions. The
reasons are as follows. Since the rigorous upper bounds for
$m(h,T)$ have been known for {\it small} $h$ by using Bogoliubov's
inequality, namely $|m(h,T)| \leq (const/T^{2/3}) |h|^{1/3}$
for one dimension,
$|m(h,T)| \leq (const/T^{1/2}) (1/|\ln|h||^{1/2})$
for two dimensions\cite{ghosh}, and $|m(h,T)| \leq (const/T^{1/2})$
for three dimensions, and in addition, the magnetic susceptibility for $U_{x
}<0$
has also been bounded by a constant in arbitrary
dimensions in Ref.\cite{kubo}, we observe that
$f(h \rightarrow 0^+,\rho,T)$ could be finite or could be divergent
but in latter case the speed of divergence must be slower than
that of $[\tanh (h \rightarrow 0^+)]^{-1} \rightarrow \infty$ in one and two
dimensions, and in three dimensions for $U_{x}<0$ as well,
otherwise it contradicts
the consequences of these upper bounds. On the other hand,
as can be seen, there are still some
rooms where $f(h \rightarrow 0^+,\rho,T)$
might be divergent for $U_{x}>0$ such that $m(h \rightarrow 0^+, T>0)$
could sustain a finite nonzero constant in this situation.
Although we have the condition (\ref{f=1}) which
is valid for any dimensions and independent of $U_{x}$,
we still can not preclude such a
possibility  from the present formalism due to the $\rho$-dependence of
$f(h,\rho,T)$.

Therefore, any solution for the function $f(h,\rho,T)$ should comply
simultaneously with the properties given in the items 1-5.
Considering these conditions obeyed by the function $f(h, \rho, T)$,
one may check that the solutions given by Eqs. (\ref{g-solu}) and (\ref{m-f})
indeed satisfy all conditions imposed on the functions ${\cal G}(h, T)$ and
$m(h,T)$ self-consistently.

Now, let us present, as an example, a possible
realization for the function $f(h,\rho,T)$, which is a well-known result
for limiting cases in the system.
Such a solution for the function $f(h,\rho,T)$ takes the simplest form,
namely,
\begin{eqnarray}
 f(h,\rho,T) = \rho,
\label{f=rho}
\end{eqnarray}
which gives rise to the following ferromagnetic
magnetization per site, $m(h,T)$, in the Hubbard model at
finite temperatures:
\begin{eqnarray}
m (h, T) = \frac{\rho}{2} \tanh (\beta h).
\label{m-f=rho}
\end{eqnarray}
It is easy to verify that this special solution
satisfies all conditions imposed on the functions $f(h,\rho, T)$,
${\cal G}(h,T)$ and
$m(h,T)$ self-consistently. As can be easily seen, it is nothing but
the result for the non-interacting system ($U_x =0$), as well as the result
for the system at half-filling, as obtained by S\"ut\H o \cite{suto}.
It leads to
the magnetic susceptibility
obeying the Curie's law, and the fluctuations of the spin-spin correlation
function in the presence of the applied field
vanishing in the thermodynamic limit. We would like to point
out here that we still can not determine if the form of $m(h,T)$ given in
Eq.(\ref{m-f=rho}) pertains to the one- and two-dimensional systems
with arbitrary $U_{x}$, and to the three-dimensional system with $U_x < 0$
from the present formalism,
although we are aware of the fact that no spontaneous ferromagnetic LRO
appears in these cases at finite temperatures.


Several remarks are collected here.

(i) As one may notice that, apart from the case at half-filling,
we can not, only from the
special solution given by Eq. (\ref{m-f=rho}), draw the conclusion that no
spontaneous ferromagnetic LRO emerges in the three-dimensional
repulsive Hubbard
model at finite temperatures, because there might be
other possible solutions  which could lead to the
exhibition of the ferromagnetic LRO at finite temperatures in some
parameter regimes. Since we
use only the symmetry properties and known exact results but not
invoking any detail information of the Hamiltonian, we can not at the moment
affirm what solution is more physically realistic in the parameter regimes w
hich
are currently interesting. It is conceivable that resolving this
problem the explicit form of the Hamiltonian (\ref{hamil}) or numerical
works ought to be
involved in the analyses. However, we expect that our
exact result [Eq. (\ref{m-f})] could shed some useful light on this issue.

(ii) We should stress that
since the
Hubbard model can be mapped onto the t-J model in the limit
of $U_x \gg t$, while the t-J model at half-filling reduces to
the antiferromagnetic Heisenberg model which may exhibit LRO at
finite temperatures in dimensions $d \geq 3$, the result [e.g. Eq.
(\ref{m-f}) or Eq.(\ref{m-f=rho}) ] is not
in conflict with this consensus.
Actually, people can expect that the antiferromagnetic LRO would occur in
the half-filled, repulsive Hubbard model at positive temperatures in three
dimensions\cite{lieb0}.

(iii) We believe that the present result is
suitable for one, two and three dimensions. It is still uncertain
whether or not the method used in this paper can be extended to
either the
Hubbard model in infinite dimensions\cite{mv},  or the Hubbard models with
artificially degenerate bands on decorated lattices\cite{mietasa},
or the spin systems which possess spin SU(2) symmetry, like the
ferromagnetic Heisenberg model, and so on,  because the additional symmetries
or restrictions in these systems, if they exist,  might
violate this method in some ways. This question remains open.

(iv) The present result holds even if we add an arbitrary one body
potential (real), e.g. $V = \sum_{x \in \Lambda} V_x (n_{x \uparrow}
+ n_{x\downarrow})$, into the Hamiltonian [Eq. (\ref{hamil})].

(v) We like to mention that
although we can not prove the uniqueness of the solutions
for ${\cal G}(h, T)$ and $m(h,T)$ at the moment, the other
solutions, if they exist, might have similar forms as those
given in this paper due to the restrictions imposed on the
two functions\cite{schad}. Despite all this, the present result
at least is {\it one} of the exact results (if any) of the model.
Moreover, our derivation, without invoking any {\it a priori} assumption,
is somewhat general.

(vi) It can be seen that any further solutions, exact or numerical, of
$m(h,T>0)$ in the Hubbard model should comply with the fundamental
equations and results discussed in this paper, otherwise they would
conflict some symmetry properties of the model.

In summary, by  means of some symmetry properties of relevant Hamiltonian,
 two fundamental relations
between the ferromagnetic magnetization and a spin correlation
function, which can be viewed as a kind of Ward-Takahashi identities,
 are derived, which allows us, combining some known exact results,
to discuss symmetry properties for the
ferromagnetic magnetization per site as a function of the applied field in the
$d (=1,2,3)$-dimensional Hubbard model
at finite temperatures.  The results hold true for both
repulsive and attractive on-site Coulomb interactions, and for
arbitrary electron fillings. As an example a special solution is also
analyzed. The properties discussed in this paper
could be helpful for further understanding the magnetic aspects of
the Hubbard model at finite temperatures.


One of the authors (G.S.)
is grateful to the Department of Applied Physics,
Science University of Tokyo, for the warm hospitality, and to the Nishina
foundation for support. This work has also been supported by the CREST
(Core Research for Evolutional Science and Technology) of the Japan
Science and Technology Corporation (JST).

\end{document}